\documentclass[aps,twocolumn,amsmath,amssymb,nofootinbib,superscriptaddress,prl]{revtex4-1}

\usepackage[colorlinks]{hyperref}
\usepackage{subfigure}
\usepackage{amsthm,amsmath,amssymb}
\usepackage{mathrsfs}
\usepackage{braket}
\usepackage{graphicx}
\usepackage{dcolumn}
\usepackage{bm}
\usepackage[noend]{algpseudocode}
\usepackage{algorithmicx,algorithm}
\usepackage{color}
\definecolor{Red}{rgb}{1,0,0}

\def\poly{\operatorname{poly}}
\def\tdqft{\operatorname{QFT2D}}
\def\odqft{\operatorname{QFT1D}}
\def\filter{\operatorname{filter}}
\def\fft{\operatorname{fft}}

\begin{document}

\title{Achieving quantum advantages for image filtering}

\author{Zidong Cui}
\affiliation{Institute of Fundamental and Frontier Sciences, University of Electronic Science and Technology of China, Chengdu, 610051, China}

\author{Shan Jin}
\affiliation{Institute of Fundamental and Frontier Sciences, University of Electronic Science and Technology of China, Chengdu, 610051, China}

\author{Akira Sone}
\affiliation{Department of Physics, University of Massachusetts, Boston, MA 02125, USA}

\author{Xiaoting Wang}
\email{xiaoting@uestc.edu.cn}
\affiliation{Institute of Fundamental and Frontier Sciences, University of Electronic Science and Technology of China, Chengdu, 610051, China}

\date{\today}

\begin{abstract}

Image processing is a fascinating field for exploring quantum algorithms. However, achieving quantum speedups turns out to be a significant challenge. In this work, we focus on image filtering to identify a class of images that can achieve a substantial speedup. We show that for images with efficient encoding and a lower bound on the signal-to-noise ratio, a quantum filtering algorithm can be constructed with a polynomial complexity in terms of the qubit number. Our algorithm combines the quantum Fourier transform with the amplitude amplification technique. To demonstrate the advantages of our approach, we apply it to three typical filtering problems. We highlight the importance of efficient encoding by illustrating that for images that cannot be efficiently encoded, the quantum advantage will diminish. Our work provides insights into the types of images that can achieve a substantial quantum speedup.
\end{abstract}


\maketitle

\textit{Introduction} -- Quantum Fourier transform (QFT) is a key component in the quantum phase estimation (QPE) circuit, which is critical in constructing Shor's and HHL algorithms~\cite{shor,hhl}. The QFT confers an exponential advantage over its classical counterpart, making it an indispensable tool for achieving quantum speedup. So far, the types of quantum circuits with a quantum speedup are rare, and typical quantum algorithms that offer quantum advantage essentially rely on either the QPE~\cite{shor} or Grover's search~\cite{grover}. Therefore, any alternative methods for achieving quantum advantage would be of great value in advancing quantum algorithms. In this work, we hope to investigate other applications of QFT where it functions independently in a quantum circuit rather than as a component of the QPE. It turns out that QFT has various applications in quantum image processing that aims at solving image processing problems in a more efficient way than classical computing~\cite{stroeIMg,store_image_Bose,flexiR,latorre2005,q_jpeg2023,PhysRevA.67.062311,Qsobel,yan2017quantum}. Quantum image processing covers a wide range of image problems, from image encoding and retrieval~\cite{stroeIMg,flexiR}, image compression~\cite{store_image_Bose,latorre2005,q_jpeg2023}, to pattern recognition~\cite{PhysRevA.67.062311} and edge extraction~\cite{Qsobel}. Among these applications, image filtering in the frequency domain serves as a pre-processing technique to modify or enhance the features of an image. For example, a high-pass filter can enhance the high-frequency components of an image while suppressing the low-frequency components. Such filtering process usually includes three steps: (1) converting the original image from the spatial domain to the frequency domain through the 2D Fourier transform, (2) constructing a filter to cutoff the unwanted frequency components, and (3) converting the filtered image back to the spatial domain through the inverse 2D Fourier transform. Among them, the essential trick is to design the appropriate cutoff for filtering in Step (2). For instance, in the case of a high-pass filter, Step (2) aims to design a cutoff frequency that attenuates components with frequency values below the cutoff, while allowing those above to pass through, leaving the output image with enhanced details and sharper edges compared to the original image. Similarly, the quantum image filtering (QImF) algorithm contains the same three steps~\cite{QIF}. Note that both Step (1) and Step (3) can be implemented through QFT with an exponential speedup, and therefore the advantage of a QImF algorithm mainly depends on the efficiency of the filtering process in Step (2). Unfortunately, in existing QImF algorithms, efficient circuit construction of Step (2) is not guaranteed, which could hinder the potential quantum speedup. It turns out that the efficiency of the quantum filtering process is closely related to the efficiency of encoding the classical image into a quantum state. Hence, here we propose a QImF algorithm with a guaranteed quantum advantage for a subgroup of image problems, by implementing Step (2) using Grover's amplitude amplification; in general, how to achieve quantum speedup for a generic image filtering problem remains open for future research.

\textit{Encoding the image as state amplitudes} -- As a preliminary step, before implementing the QImF process, the classical image must be encoded into a quantum state. Several quantum encoding methods have been proposed~\cite{stroeIMg,flexiR,encode1,encode2,encode3}; in this work, we adopt the amplitude encoding scheme, which encodes the image pixels as state amplitudes. Such encoding method works well with QFT since the outcome of the discrete Fourier transform is stored as state amplitudes in QFT. Specifically, for a digital image represented by a pixel matrix $F=(F(i,j))$, $0\le i\le N_1-1$, $0\le j\le N_2-1$, where $F(i,j)$ denotes the pixel value at the $(i,j)$-th position and takes a discrete value from $[0,1]$. Without loss of generality, we assume $N_1=2^{n_1}$, $N_2=2^{n_2}$. Then $F$ can be encoded as the amplitudes of an 
$n=(n_1+n_2)$-qubit system with state $\ket{s}=\sum_{k=0}^{2^n-1}s_{k}\ket{k}=1/\sqrt{M}\sum_{i,j}F(i,j)\ket{i,j}$, where $M$ is the normalization factor, $s_k=F(i,j)/\sqrt{M}$, and the basis $\{\ket{k}\}$ is one-to-one correspondent with the tensor-product basis $\{\ket{i,j}\equiv\ket{i}\ket{j}\}$, with $k=iN_2+j$. Let $U_e$ be the encoding unitary satisfying $\ket{s}=U_{e}\ket{0}$. In general, the circuit complexity of $U_e$ can be exponential in $n$, which affects the circuit complexity of the QImF algorithm. In this work, we only consider the image state $\ket{s}$ that can be efficiently generated, i.e., $U_{e}$ can be implemented by a quantum circuit with complexity $\poly(n)$. In fact, a wide range of image states $\ket{s}$ can be efficiently generated, including many vector images created using mathematical equations to determine each individual pixel. For example, the rectangular image can be efficiently generated, as shown in the Supplemental Material. In addition, we define the quantum filtering task as designing a unitary process $U_{\filter}$ such that $\ket{s'}=U_{\filter} \ket{s}$ corresponds to the filtered image $F'$ in the spatial domain. It should be noted that if one needs to read out the amplitudes of $\ket{s'}$, then state tomography is required whose measurement complexity could be very high; however, in many applications, image filtering is only a pre-processing step of other image processing tasks, and in this case there is no need to read out the amplitudes of $\ket{s'}$. Hence, in this work, a QImF algorithm is said to complete the filtering task successfully if it finally generates the desired filtered state $\ket{s'}$ in the spatial domain. 

\begin{figure} 
    \includegraphics[width=0.9\columnwidth]{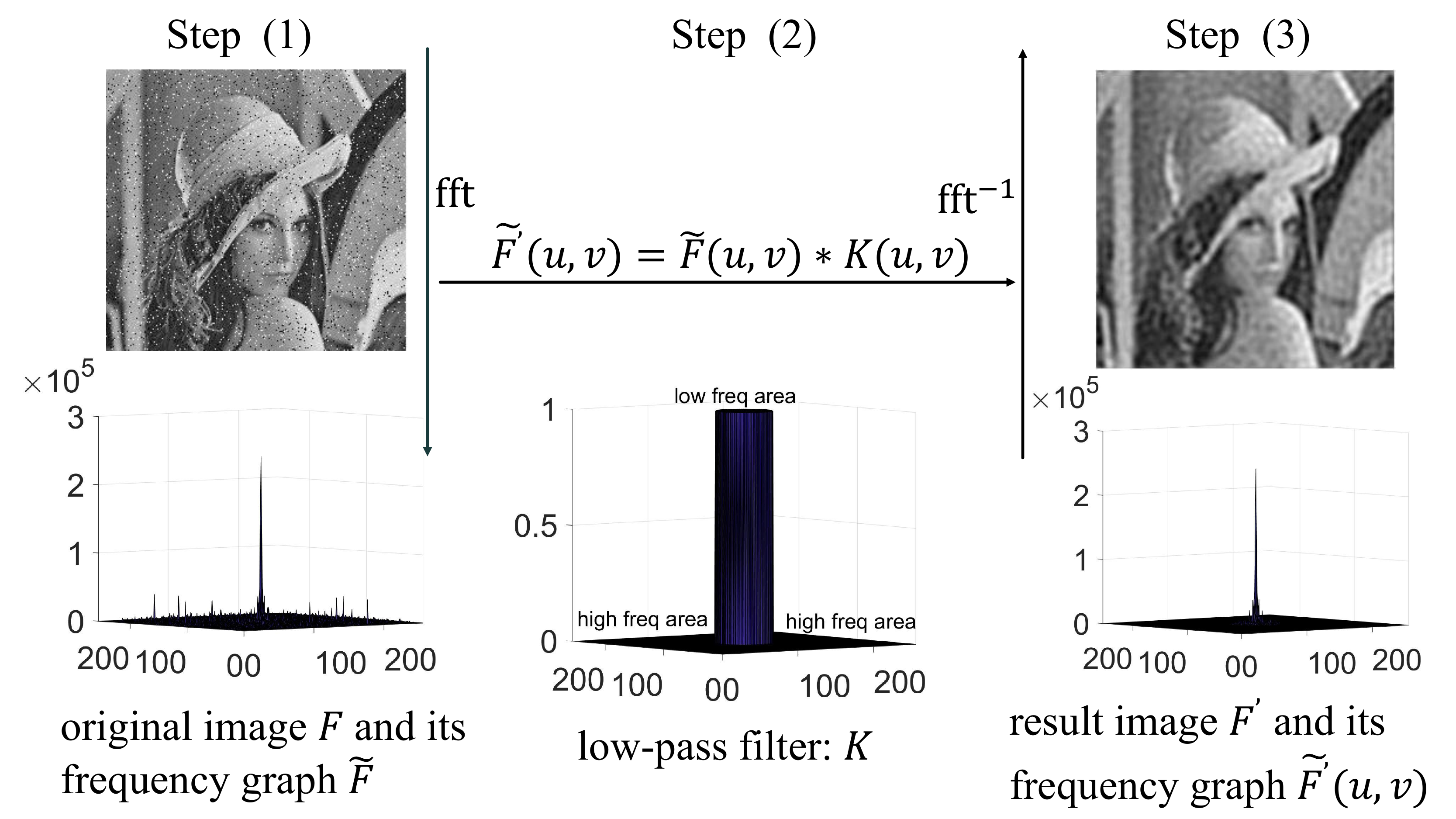}
    \caption{A low-pass filtering example to illustrate the three steps of image filtering. Steps (1) and (3) involve fast Fourier transform, and Step (2) involves pairwise multiplication. }
    \label{fig:classical_filter}
\end{figure}

\begin{figure*}
    \centering
     \includegraphics[width=0.9\textwidth]{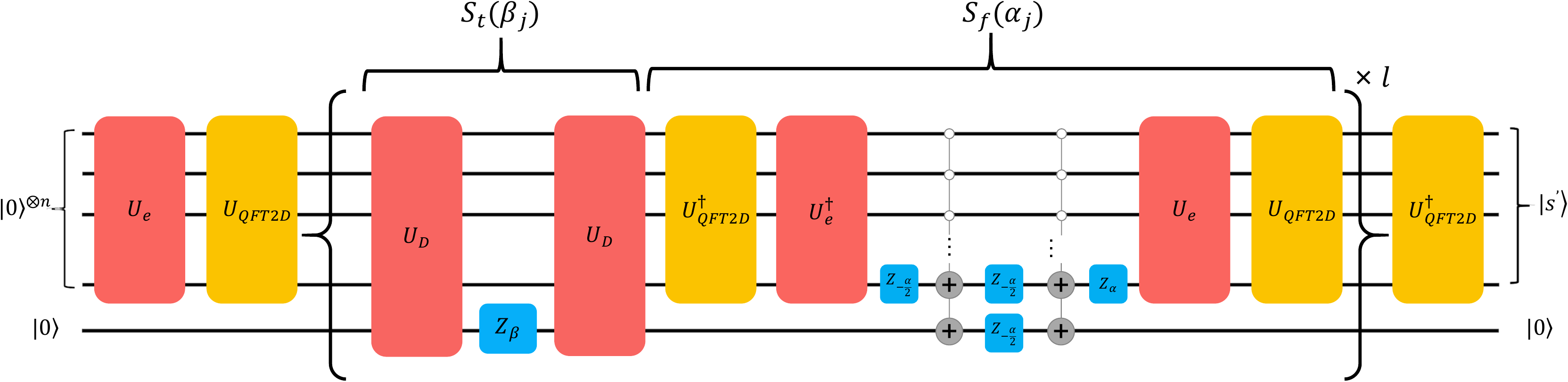}
    \caption{Quantum circuit of the quantum filtering algorithm. The first $n$ qubits form a register to store the image. The last qubit is an ancilla register used to provide the phase to flip the target bases.}
    \label{fig:circuit}
\end{figure*}


\textit{Quantum image filtering} -- We perceive $F(i,j)$ as a 2D function to describe the image $F$ in the spatial domain. Classical image filtering can be mathematically formulated into three steps, as shown in Fig.~\ref{fig:classical_filter}: (1) $F(i,j) \xrightarrow{\text{fft}}\widetilde{F}(u,v)$, (2) $\widetilde{F}'(u,v)\equiv \widetilde{F}(u,v)*K(u,v)$, and (3) $\widetilde{F}'(u,v)\xrightarrow{\text{fft}^{-1}} F'(i,j)$, where $K(u,v)$ is the filter function determined by the problem, $*$ denotes the pairwise multiplication, and $\fft$ denotes the fast Fourier transform. Likewise, after encoding $F$ into $\ket{s}$, our QImF circuit needs to implement the following three components: (i) to transform $\ket{s}$ in the spatial domain to $\ket{f}$ in the frequency domain, (ii) to construct and apply a filtering unitary to $\ket{f}$ to obtain the filtered image $\ket{f'}$, and (iii) to transform $\ket{f'}$ back to $\ket{s'}$ in the spatial domain. For an $N_1\times N_2$ image $F$ encoded in $\ket{s}$ of an $n$-qubit register, Steps (1) and (3) can be completed using the 2D QFT, which can be constructed as the tensor product of two 1D QFTs acting on the first $n_1$ qubits and the last $n_2$ qubits~\cite{FT2D1,FT2D2}:
\begin{align*}
U_{\tdqft}= U_{\odqft}^{n_1}\otimes U_{\odqft}^{n_2}
\end{align*}
where $U_{\odqft}^{n_m}(j,k)=\frac{1}{\sqrt{N_m}}e^{\frac{2\pi ijk}{N_m}}$, $m=1,2$. After applying the 2D QFT, we obtain: 
\begin{align*}
    \ket{f}\equiv U_{\tdqft}\ket{s}=\frac{1}{\sqrt{M}}\sum_{(u,v)=(0,0)}^{(N_1-1,N_2-1)}\widetilde{F}(u,v)\ket{uN_2+v}
\end{align*}
where $\widetilde{F}(u,v)$ is the amplitude at frequency $(u,v)$ and $M=\sum_{(u,v)} \widetilde{F}^*(u,v) \widetilde{F}(u,v)$ is the same normalization factor as that of $\ket{s}$, by Parseval's theorem. $M$ represents the total intensity of all frequency components. The basis state $\ket{uN_2+v}$ is one-to-one correspondence with the tensor-product basis state $\ket{u,v}\equiv\ket{u} \ket{v}$.

 
Next, we consider how to implement the pairwise multiplication in Step (2), which is straightforward on classical circuits, but becomes challenging on quantum ones. Noticing that the essence of pairwise multiplication is to amplify and suppress signals in the frequency domain, we employ an alternative process to achieve a similar effect, inspired by Grover's amplitude amplification technique~\cite{hoyer97,grover98}. By treating the desired frequency components as the target for Grover's search, we can construct Grover's iteration to selectively amplify the wanted frequency components within the amplifying region $D$ and suppress the unwanted ones outside $D$. A distance function can be used to describe $D$ by requiring the filter to amplify the frequency components based on their distance from a specific frequency point, such as the zero frequency. As there are four reference points corresponding to the zero frequency, $(0,0)$, $(0,N_2)$, $(N_1,0)$ and $(N_1,N_2)$, the amplifying region $D$ can be characterized by the difference between $(u, v)$ and its nearest zero frequency point:
\begin{align*}
d_0(u,v)&\equiv \min_{u_0\in\{0,N_1\},v_0\in\{0,N_2\}} \sqrt{(u-u_0)^2+(v-v_0)^2}\\
D&\equiv \left\{(u,v)|d_1\le d_0(u,v) \le d_2\right\}
\end{align*}
Thus, $D$ is composed of four quarter-annulus regions centered at the four zero reference points. Based on $D$, the target state $\ket{t}$ for Grover's search is defined as:
\begin{align}
    \ket{t}\equiv\frac{1}{\sqrt{M_{D}}}\sum_{(u,v)\in D}\widetilde{F}(u,v)\ket{uN_2+v}
\label{notion:t}
\end{align}
where $M_{D}\equiv \sum_{(u,v)\in D} \widetilde{F}^*(u,v) \widetilde{F}(u,v)<M$ is the normalization factor of $\ket{t}$, representing the total intensity in the amplifying region $D$. Accordingly, the quantum oracle $U_D$ can be constructed, satisfying:
    \begin{align}
        U_D\ket{uN_2+v}\ket{b} = 
        \begin{cases}
        \ket{uN_2+v}\ket{b\oplus 1}, & \text{if } (u,v)\in D \\
        \ket{(uN_2+v}\ket{b}, & \text{if } (u,v)\notin D
        \end{cases}
\end{align}
where $\ket{b}$ is the ancilla state, $b=0,1$. The role of $U_D$ is to judge whether $(u,v)\in D$ for each $(u,v)$. Note that several improved search protocols have been proposed~\cite{grover_fixed,YLC}, besides the standard approach of Grover's search~\cite{grover}. In this work, we will adopt YLC's fixed-point search algorithm, which prevents the overcooking problem as in the standard approach while maintaining the quadratic speedup~\cite{YLC}. Specifically, based on $U_D$, we construct the YLC's iteration sequence as follows: 
\begin{subequations}
\begin{align}
    &S_{f}(\alpha_{j})=I-(1-e^{-i\alpha_{j}})\ket{f}\bra{f}\\
    &S_{t}(\beta_{j})=I-(1-e^{i\beta_{j}})\ket{t}\bra{t}\\
    &U_{I}=\prod_{j=1}^{l}G(\alpha_{j},\beta_{j})=\prod_{j=1}^{l}-S_{f}(\alpha_{j})S_{t}(\beta_{j})
\end{align}
\end{subequations}
where $U_{I}$ is the iteration sequence composed of $l$ Grover's iteration $G(\alpha_{j},\beta_{j})\equiv -S_{f}(\alpha_{j})S_{t}(\beta_{j})$ formed by two rotations $S_t$ and $S_f$ with rotation angles $\{\alpha_j,\beta_k\}$ chosen as:
\begin{subequations}
\begin{align}
    &\alpha_{j}=-\beta_{l-j+i}=2 \rm{arccot} \left(\tan(2\pi j/L)\sqrt{1-\gamma^{2}}\right)\\
    &\gamma^{-1}=T_{1/L}(1/\delta)=\cos\left[L \arccos(1/\delta)\right]\\
    &L\equiv 2l+1\geq \frac{\log(2/\delta)}{\sqrt{\lambda}},
\label{para}
\end{align}
\end{subequations}
where $T_{L}(x)\equiv \cos\big(L\arccos(x)\big)$ is the $L$-th Chebyshev polynomial of the first kind~\cite{YLC}. Here, $\delta$ is an error threshold chosen for the filtering problem and we require that at the end of $U_I$ the success probability is larger than $1-\delta^2$. $l$ and $L$ are used to measure the sequence length of $U_I$. $\lambda\equiv |\braket{f|t}|^2$ describes the overlap between $\ket{t}$ and $\ket{f}$. For given $\delta$ and $\lambda$, the minimum $l$ is calculated, and the entire filtering circuit is constructed as shown in Fig.~\ref{fig:circuit}. When passing through the circuit $U_I$, the quantum state is always in the subspace spanned by $\{\ket{t}, \ket{\bar{t}}\}$, where $\ket{\bar{t}}=\frac{1}{\sqrt{M'_{D}}}\sum_{(u,v)\notin D}\widetilde{F}(u,v)\ket{uN_2+v}$, and $\ket t\perp \ket{\bar{t}}$. We have $M'_{D}=\sum_{(u,v)\notin D} \widetilde{F}^*(u,v) \widetilde{F}(u,v)=M-M_D$, representing the total intensity outside $D$. Hence, $S_t(\beta)$ can be implemented by $U_DZ_\beta U_D$ with $U_D=U_D^\dag$, where $Z_\beta=\text{diag}(e^{-i\beta/2},e^{i\beta/2})$ is the phase gate. Similarly, the rotation $S_f(\alpha)$ with respect to $\ket{f}=U_{\tdqft}U_e\ket{0}$ can be implemented by $U_e$, $U_{\tdqft}$, and $Z_\alpha$. Since $U_e$ is required in constructing $S_f(\alpha)$, the efficiency of generating $U_e$ is important for the efficiency of $U_I$. Finally, at the end of $U_I$, another $U_{\tdqft}^\dag$ is applied to transform the filtered state $\ket{f'}$ to $\ket{s'}$ in the spatial domain. In practice, by choosing appropriate values of $D$, $\delta$ and $l$, the output $\ket{s'}$ will correspond to a satisfying filtered image. 

Note that the efficiency of $U_e$ only guarantees the efficiency of a single Grover iteration. To achieve the efficiency of the entire amplitude amplification circuit, we also need to consider the query complexity characterized by the iteration number $l$, which is then determined by $\lambda$:
\begin{align*}
\lambda= |\braket{f|t}|^2=\frac{M_D}{M}=\frac{M_D}{M_D+M'_D}
\end{align*}
Here, $\lambda$ denotes the intensity portion of the frequency components we want to retain during filtering. When we apply a filter to an image, the aim is to amplify the wanted components while suppressing the unwanted components. Here, for a given filtering task, we call the wanted signal components in $D$ as the \emph{signal} components, and the unwanted components outside $D$ as the \emph{noise} components. Then the signal-to-noise ratio of the image is defined as $\lambda'\equiv M_D/M'_D$. 
In theory, as the image size $N=N_1N_2$ increases, if the value of $\lambda'$ becomes too small, then $\lambda$ will also become too small, leading to a long and inefficient iteration sequence. However, in practice, if $\lambda'$ is too small, then the filter may not be able to distinguish between signal and noise, and may end up filtering out useful information along with the noise. In other words, if $\lambda'$ or $\lambda$ is too small, the filtered image may not be able to provide any useful information. Therefore, it generally requires that all images to be filtered have sufficient contrast and detail to facilitate effective filtering, i.e., the value of $\lambda$ for these images should exceed a certain threshold $\lambda_0>0$. Under the assumption $\lambda\ge \lambda_0$, the iteration sequence length $l$ becomes independent of $N$, allowing the entire circuit to operate efficiently. Note that $\lambda$ being lower-bounded is equivalent to $\lambda'$ being lower-bounded. 

\textit{Complexity analysis} -- The above analysis shows that, under a given filtering task with a chosen $D$, our QImF circuit in Fig.~\ref{fig:circuit} is efficient for the images with efficient encoding and a lower-bounded $\lambda$. Specifically, as the image size $N$ increases, efficient encoding guarantees that $U_e$ has a complexity of $O(\poly(\log N))$. Next, since the classical circuit to judge whether $(u,v) \in D$ for a given $(u,v)$ has a complexity of $O(\poly(\log N))$, the gate complexity of constructing $U_D$ is also $O(\poly(\log N))$. Hence the complexity of each Grover's iteration is $O(\poly(\log N))$. Under the further assumption that $\lambda\ge \lambda_0$, the iteration sequence length $l$ (i.e. the query complexity) is independent of $N$, and hence the total gate complexity of the circuit in Fig.~\ref{fig:circuit} is $O(\poly(\log N))$. In contrast, performing the same amplification process on a classical computer requires setting $F(u,v)$ to $0$ for $(u,v)\in D$ and preserving it for all other values, and this necessitates using the classical oracle to determine whether $(u,v)\in D$ for every $(u,v)$, resulting in an $O(N)$ complexity. If we do not implement the same amplification process, but a different pairwise multiplication, such calculation also has an $O(N)$ complexity. Hence, the total complexity of the classical image filtering is $O\left(\poly(N)+N\log N\right)$, where $O(N\log N)$ comes from the complexity of the fast Fourier transform. Such unique feature of images with efficient encoding and a lower bound on $\lambda$ makes our QImF algorithm highly efficient for filtering tasks. However, it is important to note that the efficient-encoding assumption is crucial, since for images that do not have an efficient encoding, the quantum speedup may completely disappear.

\textit{Applications} -- Next, we demonstrate our QImF algorithm through three filtering problems. 

\begin{figure}
    \centering
    \includegraphics[width=0.8\columnwidth]{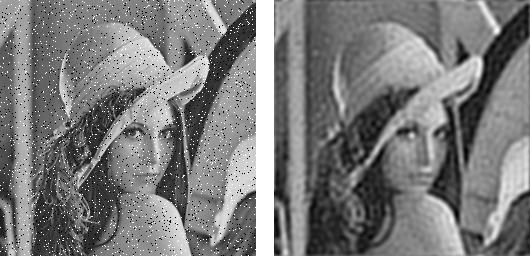}\\
\caption{Effect of low-pass filtering under the proposed QImF algorithm. Impulse noise is added into the portrait image~\cite{lena} (left), and is significantly reduced in the filtered image (right). } 
    \label{fig:low_pass}
\end{figure}

As the first application, low-pass filtering is a technique in computer vision for image smoothing and noise reduction~\cite{low_high_back1}. Here, we consider a $256\times 256$ portrait image with salt-and-pepper noise shown in the left-hand side of Fig.~\ref{fig:low_pass}. Such impulse noise is mostly located in the high-frequency range, and a low-pass filter is introduced to remove it. Specifically, we choose the amplifying region $D=\left\{(u,v)|0=d_1\le d_0(u,v)\le d_2=35 \right\}$.
The values of $d_1$ and $d_2$ are typically determined through observation and experience, often involving a trial-and-error process.  
Then $U_D$ and $U_I$ can be constructed as shown in Fig.~\ref{fig:circuit}. By directly observing the image, one can easily tell the portrait portion from the noise, implying that the signal-to-noise ratio $\lambda'=M_D/M_D'$ is pretty large, and the value of $\lambda$ should be close to $1$ (accurate calculation gives $\lambda =0.88$). Hence, for the given error threshold $\delta=0.01$, a short amplitude-amplification sequence (e.g., $l=1$) is already good enough to complete the filtering. Once the value of $l$ is chosen, we can calculate the angles $(\alpha_j,\beta_j)$ for each iteration $j$, and construct the corresponding low-pass-filtering circuit in Fig.~\ref{fig:circuit}. Simulations suggest that, an iteration sequence with $l=1$ will produce a filtered image with the impulse noise significantly reduced, as shown in Fig.~\ref{fig:low_pass}. Note that the filtered image is generated using the amplitudes of the filtered image state $\ket{s'}$ calculated through simulation. 

\begin{figure}
\centering
\includegraphics[width=0.8\columnwidth]{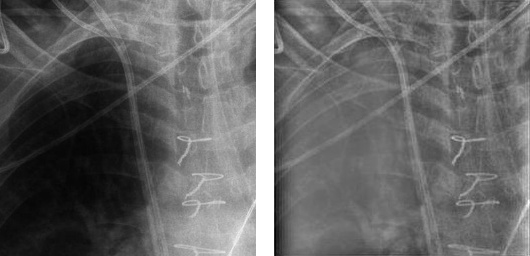}
\caption{Effect of high-pass filtering under the proposed QImF algorithm. Details in the dark region of the chest image~\cite{x_ray} (left) are clearly displayed in the filtered image (right).}
\label{fig:high_pass}
\end{figure}

As the second application, high-pass filtering is a technique that extracts and emphasizes edge information in images, and has extensive application in medical imaging~\cite{high_back,high_back2,high_back3,low_high_back1}. Here, we apply our QImF algorithm to a chest X-ray image for high-pass filtering, as shown in Fig.~\ref{fig:high_pass}. The desired frequency components that need to be preserved are in the high-frequency range, and we choose $D=\left\{(u,v)|5=d_1\le d_0(u,v))\le d_2=60\right\}$.
For $\delta=0.01$, numerical simulation suggests that Grover's iteration sequence with $l=5$ is sufficient to produce a satisfying filtered image: the black region of the original image is clearly displayed in the filtered image, which is beneficial for subsequent medical diagnosis. 


\begin{figure}
    \centering
    \includegraphics[width=0.8\columnwidth]{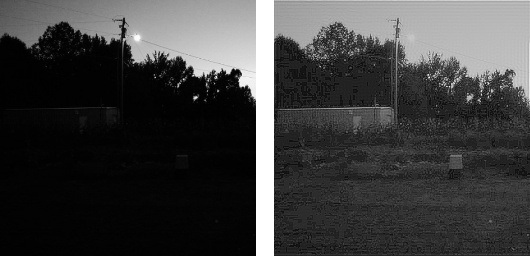}
    \caption{Effect of homomorphic filtering under the proposed QImF algorithm. Compared to the original image~\cite{homoIm} (left), the filtered image shows a significant improvement in the balance of brightness and contrast (right).}
    \label{fig:homo}
\end{figure}

As the third application, homomorphic filtering is a powerful technique for enhancing images by simultaneously increasing contrast and standardizing brightness. The homomorphic filtering is applied to an image in its illumination-reflection format, where the frequency components are divided into the illumination part and the reflection part~\cite{homo_back1,homo_back2}. The goal is to balance the ratio of the two parts to achieve better brightness and contrast. Here, we apply our QImF algorithm to a scenery photo image with flawed lighting~\cite{homoIm} for homomorphic filtering, as shown in Fig.~\ref{fig:homo}. The original image is first converted into its illumination-reflection format, which will then be encoded into the input state $\ket{s}=U_e\ket{0}$ in Fig.~\ref{fig:circuit}. The amplifying region homomorphic filtering is chosen as $D=\left\{(u,v)|75=d_1\le d_0(u,v)\le d_2=90 )\right\}$.
For $\delta=0.01$, simulations suggest that Grover's iteration sequence with $l=2$ is sufficient to produce a desired filtered image in Fig.~\ref{fig:homo}, demonstrating a significant improvement in the balance of brightness and contrast.


\textit{Discussion} -- We have identified a subset of images for which an efficient QImF algorithm can be constructed with complexity $O(\poly(\log(N)))$. Note that efficient encoding and a lower-bound on the signal-to-noise ratio are sufficient conditions to achieve the quantum speedup. Restricting our focus to this specific subset of images is crucial in establishing the quantum advantage, given the difficulty of proving the quantum speedup for all images. This is similar in importance to the $s$-sparse condition for the HHL algorithm~\cite{hhl} or the low-rank condition for the QPCA algorithm~\cite{qpca}. For future work, it is worthwhile to study how these conditions can be relaxed or modified.

\begin{acknowledgments}
Z. C., S. J., and X. W. are supported by the National Natural Science Foundation of China under Grant No. 92265208 and the National Key R\&D Program of China under Grant No. 2018YFA0306703. A. S. gratefully acknowledges startup funding supported by the University of Massachusetts, Boston. 
\end{acknowledgments}


\bibliography{QImF.bib}


\end{document}


\title{Supplemental material for ``Achieving quantum advantages for image filtering"}

\author{Zidong Cui}
\affiliation{Institute of Fundamental and Frontier Sciences, University of Electronic Science and Technology of China, Chengdu, 610051, China}

\author{Shan Jin}
\affiliation{Institute of Fundamental and Frontier Sciences, University of Electronic Science and Technology of China, Chengdu, 610051, China}

\author{Akira Sone}
\affiliation{Department of Physics, University of Massachusetts, Boston, MA 02125, USA}

\author{Xiaoting Wang}
\affiliation{Institute of Fundamental and Frontier Sciences, University of Electronic Science and Technology of China, Chengdu, 610051, China}

\date{\today}

\maketitle

\section{An example of image states that can be efficiently generated}

\begin{figure}
    \centering
    \includegraphics[width=3.4in]{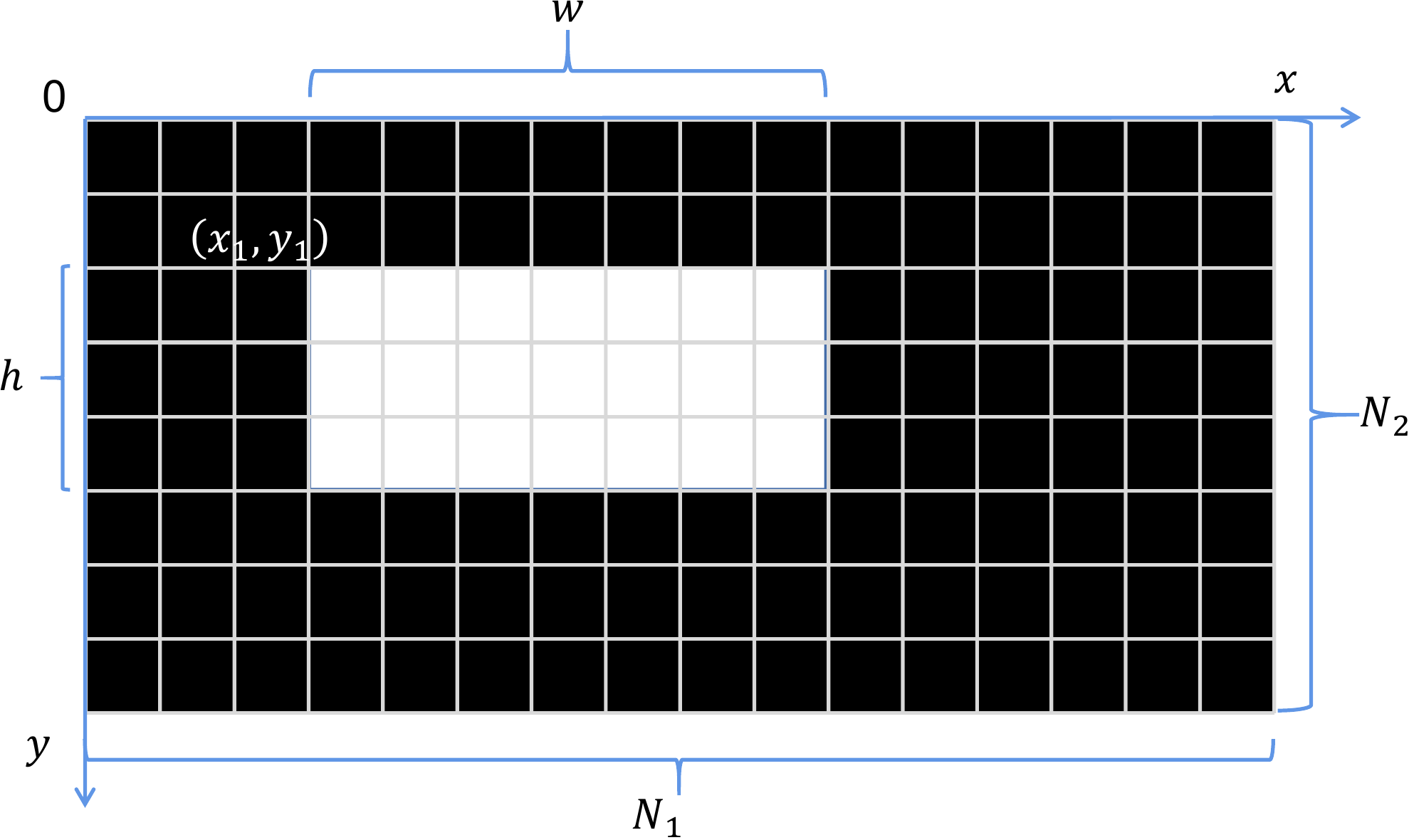}
    \caption{A $N_1\times N_2$ image $F$ containing a $w\times h$ rectangle whose top-left vertex is located at $(x_1,y_1)$. All pixels within the white rectangle takes pixel value of $p=255$; all pixels outside the rectangle have $p=0$. }
    \label{fig:rec_image}
\end{figure}

A wide range of images can be efficiently encoded as quantum states. To demonstrate this, we take a rectangle vector image $F$ as an example to illustrate how to efficiently construct $U_e$ that maps $\ket{0}$ into the image state of $F$. We assume the black-white image $F$ has dimensions $N_1\times N_2$ and contains a rectangle sized $w\times h$. Each pixel of $F$ is located at $(x,y)$ satisfying $x\in [0,N_1]$, $y\in [0,N_2]$, and $x$, $y$ are integers. The rectangle is oriented with its sides in parallel with the two axes and its top-left vertex located at $(x_1,y_1)$, as shown in Fig.~\ref{fig:rec_image}. In computer vision, black-white images are typically represented as grayscale images, where each pixel has a single intensity value ranging from $p=0$ (black) to $p=255$ (white). In this example, we assume the rectangle region is white ($p=255$) and the background is black ($p=0$). Hence, under the amplitude encoding scheme, the corresponding image state $\ket{s}$ of $F$ can be written as an equal superposition of basis states $\ket{j,k}$ within the rectangle region:
\begin{align}
\ket{s}=\frac{1}{\sqrt{M}}\sum_{x_1\le j\le x_1+w,\,y_1\le k\le y_1+h}\ket{j,k},
\label{s_image_state}
\end{align}
where $M$ is the normalization factor. In the following, we will explain how to construct $U_e$ such that $\ket{s}=U_e\ket{0}$. The entire encoding process can be divided into two steps:


Step 1: from the initial state $\ket{0}$, we first construct $U_C$ that can generate the equal-superposition state 
\begin{align}
\ket{s_C}=U_C\ket{0}=\frac{1}{\sqrt{M_C}}\sum_{j=0}^C\ket{j},
\label{independent_variable}
\end{align}
composed of all basis states $\ket{j}$ with $\{j\in \Z | 0\le j\le C\}$. Then we can use the same technique to generate the following two equal-superposition states of two subsystems: 
\begin{subequations}
\begin{align}
\ket{s_w}&\equiv\frac{1}{\sqrt{M_w}}\sum_{j=0}^w\ket{j}=U_w\ket{0}, \label{s_w}\\ 
\ket{s_h}&\equiv\frac{1}{\sqrt{M_h}}\sum_{k=0}^h\ket{k}=U_h\ket{0} \label{s_h}
\end{align}
\end{subequations}
Then by combining the two subsystems, we can construct the following equal-superposition state: 
\begin{align}
\ket{s_0}\equiv \ket{s_w}\ket{s_h}=\frac{1}{\sqrt{M}}\sum_{0\le j\le w,0\le k\le h}\ket{j,k}=U_w\otimes U_h \ket{0}\ket{0} ,
\label{s_0}
\end{align}
corresponding to the rectangle image with the top-left vertex located at $(0,0)$. 

\begin{figure}
    \centering
\includegraphics[width=0.7\columnwidth]{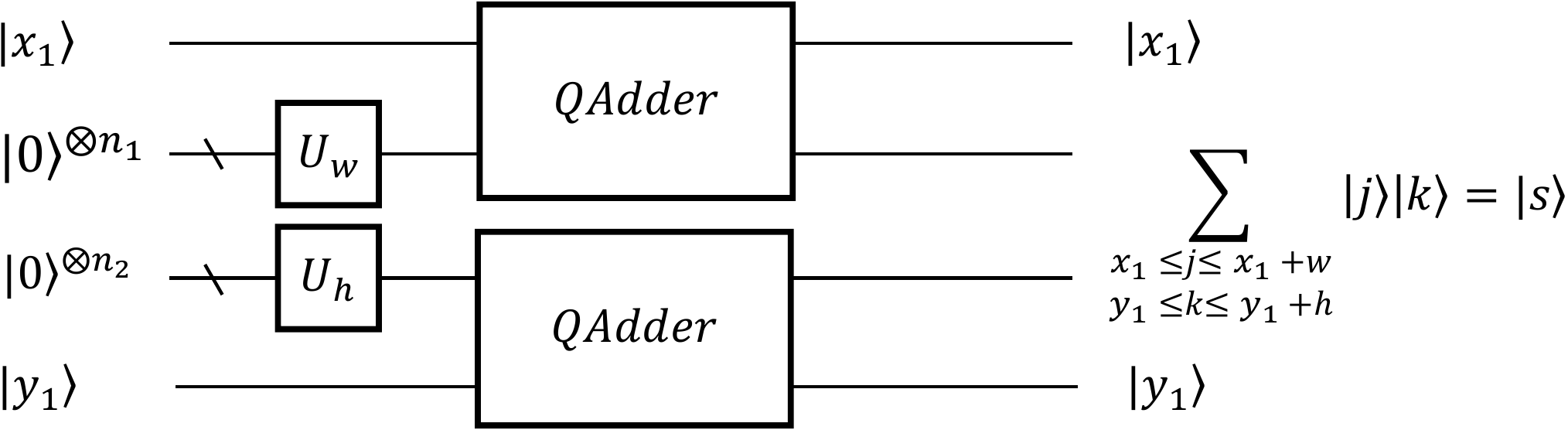}
    \caption{An efficient implementation of $U_e$ to generate the rectangle image state $\ket{s}$ in Eq.~\eqref{s_image_state}. }
    \label{circuit_buildrec}
\end{figure}

Step 2:  after applying $U_w\otimes U_h$, we can utilize two quantum addition circuits and two ancilla to translate the rectangle from $(0,0)$ to $(x_1,y_1)$, with the entire circuit shown in Fig.~\ref{circuit_buildrec}. 

\begin{figure}
    \centering
\includegraphics[width=0.7\columnwidth]{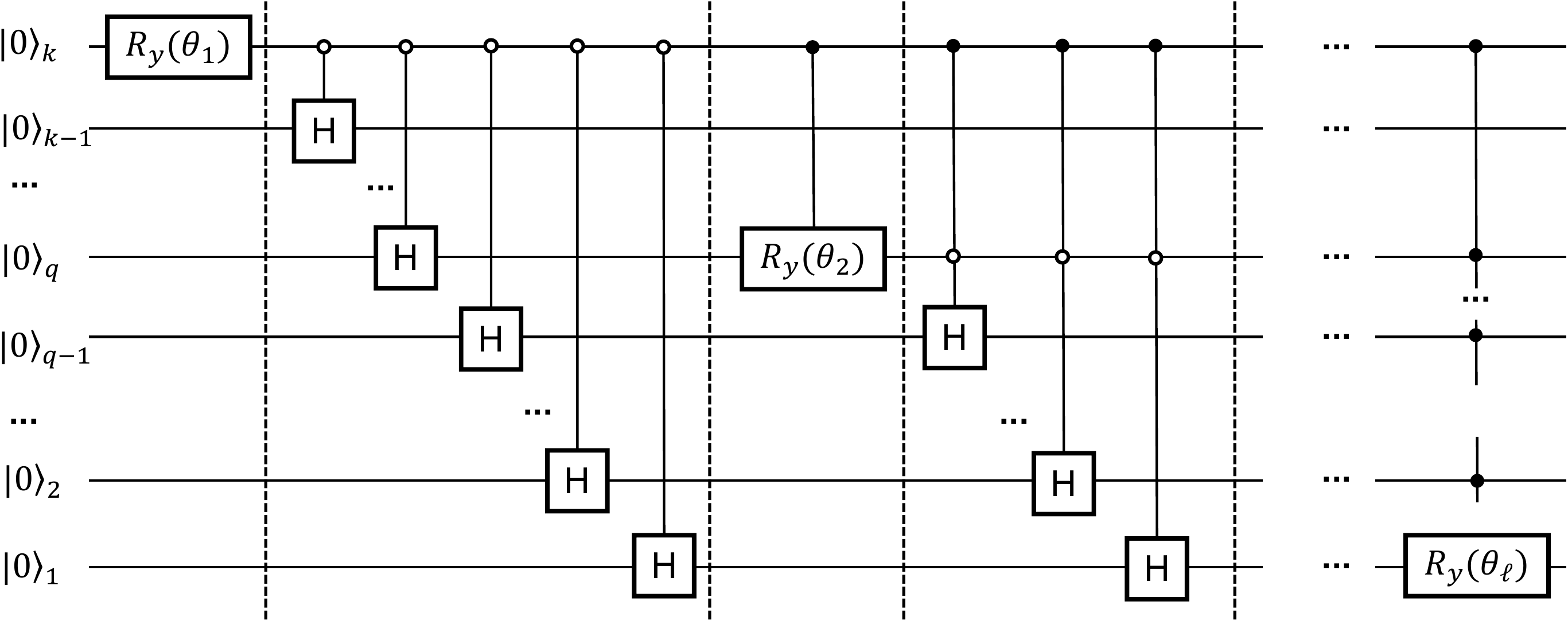}
    \caption{The quantum circuit $U_C$ to generate the equal-superposition state $\ket{s_C}=\frac{1}{\sqrt{M_C}}\sum_{j=0}^C\ket{j}$ composed of all basis states $\ket{j}$ with $\{j\in \Z | 0\le j\le C\}$.}
    \label{circuit stateInde}
\end{figure}

In the following, we denote $\Omega=\{j\in \Z | 0\le j\le C\}$ as the set of all integers within the interval $[0,C]$, where $C$ is an integer. Next, $C$ is rewritten into the binary form: $C=(c_kc_{k-1}\ldots c_2c_1)$, with $c_k=1$ as the first nonzero digit from left to right. In the binary form, $\Omega$ can be divided into two subsets: 
\begin{align*}
\Omega&=\{j\in \Z | (0_k0_{k-1}\ldots 0_20_1)\le j\le (1_kc_{k-1}\ldots c_2c_1)\}\\
&=\{ 0\le j\le (0_k1_{k-1}\ldots 1_21_1)\}\cup \{ (1_k0_{k-1}\ldots 0_20_1)\le j\le (1_kc_{k-1}\ldots c_2c_1)\}\equiv \Omega_1\cup \Omega_{2'}
\end{align*}
Let $c_q$ be the next non-zero number after $c_k$ from left to right, with $1\le q<k$. Then $\Omega$ can be further divided into: 
\begin{align*}
\Omega&=\{0\le j\le (0_k1_{k-1}\ldots 1_21_1)\}\cup \{ (1_k0_{k-1}\ldots 0_20_1)\le j\le (1_k0_{k-1}\ldots0_q1_{q-1}\ldots 1_1)\}\\
&\cup \{(1_k0_{k-1}\ldots 1_q0_{q-1}\ldots 0_1)\le j\le (1_k0_{k-1}\ldots1_qc_{q-1}\ldots c_1)\}\equiv \Omega_1\cup \Omega_2 \cup \Omega_{3'}
\end{align*}
We continue this dividing process until reaching the last non-zero digit $c_m$ in $C$ from left to right, and then we complete the last division based on $c_m$. Thus, we have divided the interval $[0,C]$ into $\ell$ subintervals, and $\Omega$ into $\ell$ subsets $\Omega_1$, $\Omega_2$,\ldots,$\Omega_\ell$, $\ell\le k\le \lfloor \log_2 C\rfloor +1 $. Denoting the cardinality of $\Omega_j$ as $M_{j}\equiv |\Omega_j|$, $j=1,\cdots,\ell$, one can check that the quantum circuit in Fig.~\ref{circuit stateInde} will generate the desired $U_C$ satisfying Eq.~(\ref{independent_variable}), where the y-rotation $R_{y}(\theta_{j})$ and the rotation angle $\theta_{j}$ are defined as:
\begin{equation}
R_{y}(\theta_{j})=
    \left(\begin{array}{cc}
        \sqrt{\frac{M_{j}}{M_{j-1}}} &-\sqrt{1-\frac{M_{j}}{M_{j-1}}}\\
        \sqrt{1-\frac{M_{j}}{M_{j-1}}} & \sqrt{\frac{M_{j}}{M_{j-1}}}
    \end{array}\right)
    \label{Ry+},\ M_{0}=C
\end{equation}

The entire encoding circuit is illustrated in Fig.~\ref{circuit_buildrec}. Without loss of generality, we assume $N_{1}=2^{n_1}$, and $N_2=2^{n_2}$. Then $n=n_1+n_2$ are required to encode the image $F$. First, similar to $U_C$, we can construct $U_w$ and $U_h$ in Eq.~\eqref{s_w} and Eq.~\eqref{s_h} to generate $\ket{s_0}$ in Eq.~\eqref{s_0}, as shown in Fig.~\ref{circuit_buildrec}. $\ket{s_0}$ represents the rectangle image with  
the top-left vertex located at $(0,0)$. Then a translation unitary can be constructed based on the quantum addition circuit, QAdder~\cite{qadder}, to shift the rectangle from $(0,0)$ to $(x_1,y_1)$, as shown in Fig.~\ref{circuit_buildrec}. One can easily check that the final output is indeed the desired encoding state $\ket{s}$ in Eq.~\eqref{s_image_state}. 

Now we check the complexity of the circuit $U_e$ in Fig.~\ref{circuit_buildrec}. Since the complexity of $U_C$ is $O(\poly(\ell))=O(\poly(\log_2 C))$, we conclude that the complexity of $U_w\otimes U_h$ is $O(\poly(n))$. On the other hand, since the classical complexity of addition is $O(\poly(n))$ for $n$-digit addition, the quantum complexity of the two QAdder's is $O(\poly(n_1))+O(\poly(n_2))=O(\poly(n))$. Hence, the total complexity of $U_e$ in Fig.~\ref{circuit_buildrec} is $O(\poly(n))$, i.e., the image state $\ket{s}$ of the rectangle image can be efficiently generated. 

\begin{figure}
    \centering
    \includegraphics[width=0.4\columnwidth]{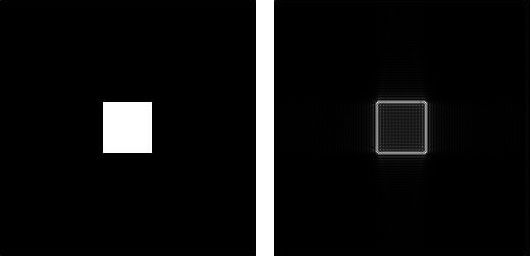}
   \caption{The simulated outcome of the filtered image state $\ket{s'}$, after the high-passing filtering circuit for the rectangle image $F$.    
The quantum high-pass filtering circuit for the rectangle image $F$. The figure on the left is the image generated by the quantum circuit in Fig.~\ref{circuit_buildrec}. We set $N_1=N_2=256$, $w=h=50$, $x_1=y_1=103$. The figure on the right is the result after high-pass filtering, with $d_1=140,\ d_2=80,\ l=15$. In this example, the proportion of the frequency of the edge of the rectangle we want to extract is small, so we use more iterations to amplify the desired frequency. However, the proportion of this frequency does not change much with the increase of image resolution, and is almost a constant value.}      
\label{fig:exampleRec}
\end{figure}

Next, we apply our QImF algorithm to extract the edges of the rectangle image through high-pass filtering. Specifically, we set $N_1=N_2=256$, $w=h=50$, $x_1=y_1=103$. We also choose the amplifying region $D$:
\begin{equation}
D=\left\{(u,v)|80=d_1\le d_0(u,v)\le d_2=140 )\right\}
\end{equation}
Then by choosing $\delta=0.01$ and the iteration sequence length parameter $l=15$, we can implement the QImF circuit and derive the final filtered image state $\ket{s'}$, as shown in Fig.~\ref{fig:exampleRec}. One can see that the filtered state $\ket{s'}$ does correspond to a satisfying filtering outcome.

\bibliography{QImF.bib}